\documentclass[aps,prb,twocolumn,superscriptaddress,showpacs]{revtex4}
\usepackage[dvips]{graphicx}
\usepackage{color}
\begin{document}
\hyphenation{VOCl TiOCl TiOBr TiOX PHOIBOS SAES Hei-sen-berg oxy-ha-li-des
pseu-do-gap mul-ti-plet quasi-par-ti-cle iso-struc-tu-ral mo-no-chro-ma-ti-zed
an-ti-fer-ro-mag-ne-tic an-ti-fer-ro-mag-ne-ti-cal-ly or-bi-tals sand-wi-ched}
\title{Electronic structure of the two-dimensional Heisenberg antiferromagnet VOCl: a multi-orbital Mott insulator}

\author{S.~Glawion}
\author{M. R.~Scholz}
\affiliation{Experimentelle Physik 4, Universit\"at W\"urzburg, 97074 W\"urzburg,
Germany}
\author{Y.-Z.~Zhang}
\author{R.~Valent\'i}
\affiliation{Institut f\"ur Theoretische Physik, Universit\"at Frankfurt, 60054
Frankfurt/Main, Germany}
\author{T.~Saha-Dasgupta}
\affiliation{SN Bose National Centre for Basic Sciences, Salt Lake City, Kolkata
700098, India}
\author{M.~Klemm}
\affiliation{Experimentalphysik II, Institut f\"ur Physik, Universit\"at Augsburg,
86135 Augsburg, Germany}
\author{J.~Hemberger}
\affiliation{Experimentalphysik II, Institut f\"ur Physik, Universit\"at Augsburg,
86135 Augsburg, Germany} \affiliation{II. Physikalisches Institut, Universit\"at zu
K\"oln, 50923 K\"oln, Germany}
\author{S.~Horn}
\affiliation{Experimentalphysik II, Institut f\"ur Physik, Universit\"at Augsburg,
86135 Augsburg, Germany}
\author{M.~Sing}
\author{R.~Claessen}
\affiliation{Experimentelle Physik 4, Universit\"at W\"urzburg, 97074 W\"urzburg,
Germany}

\date{\today}

\begin{abstract}
We have studied the electronic structure of the two-dimensional Heisenberg
antiferromagnet VOCl using photoemission spectroscopy and density functional theory
including local Coulomb repulsion. From calculated exchange integrals and the
observed energy dispersions we argue that the degree of one-dimensionality regarding
both the magnetic and electronic properties is noticeably reduced compared to the
isostructural compounds TiOCl and TiOBr. Also, our analysis provides conclusive
justification to classify VOCl as a multi-orbital Mott insulator. In contrast to the
titanium based compounds density functional theory here gives a better description
of the electronic structure. However, a quantitative account of the low-energy
features and detailed line shapes calls for further investigations including
dynamical and spatial correlations.
\end{abstract}



\maketitle

\section{Introduction}
Transition metal compounds are characterized by a plethora of many-body phenomena
which emerge from a complex interplay of charge, spin, orbital, and lattice degrees
of freedom. More recently, strong focus has been placed on the role of multi-orbital
physics, which is governed by a competition between intra-atomic Hund's rule
exchange coupling and crystal field splitting in the $d$
shell.\cite{Liebsch03,Costi07,Medici09} The simplest possible case where
multi-orbital effects may occur is in a $d^2$ configuration, as encountered, e.g.,
in trivalent vanadium compounds. A prominent example is V$_2$O$_3$, whose spin
nature, whether being $S=1$, as naively expected for a $d^2$ system and found
spectroscopically,\cite{Park00} or whether V-V pair dimerization leads to an
effective $S=1/2$ behavior,\cite{Castellani78} has been lengthly debated and
recently settled as a $S=1$ system (for details, see Ref.~\onlinecite{Poteryaev07}).

Here we study another prototypical V$^{3+}$ compound, namely VOCl. It belongs to the
class of low-dimensional transition metal oxyhalides of the form \emph{M}O\emph{X}
(\emph{M}=Ti,V,Cr,Fe; \emph{X}=Cl,Br,I). Amongst these, especially the $3d^1$
systems TiOCl and TiOBr have seen considerable interest
recently,\cite{Seidel03,Lemmens05,Hoinkis05,Krimmel06,Abel07,Kuntscher08,Zhang08} as
they also show dimerization leading to non-conventional spin-Peierls behavior. Like
its Ti counterparts, VOCl has a layered crystal structure which provides a
topologically frustrated lattice, thus being a good candidate for unusual ground
states and phase transitions. Such tendencies are enhanced due to competing one- and
two-dimensional
interactions.\cite{Rueckamp05a,Clancy07,Mastro09,Hoinkis07,Aichhorn08} Furthermore,
it has been shown that TiO\emph{X} can be electron-doped by alkali metal
intercalation into the van der Waals gaps of the crystal structure\cite{Sing09} up
to a nominal $d^{1+x}$ configuration. In some sense, the electronic structure of
VOCl can be viewed as the $x=1$ limit of such a doping series and thus provides an
interesting reference point.
\begin{figure}
\includegraphics[width=0.48\textwidth]{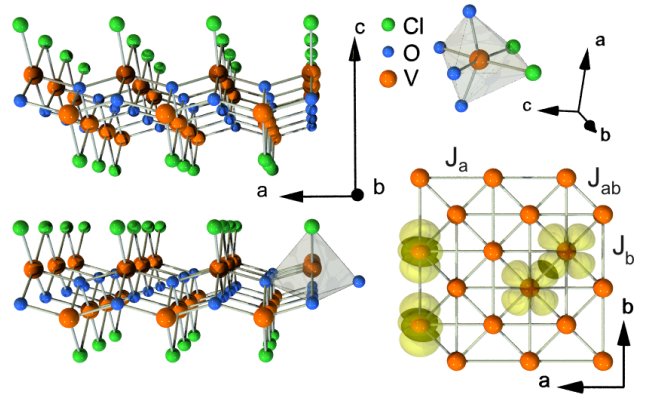}
\caption{\label{Figure1}(Color online) Left: Three-dimensional view of the VOCl
crystal structure along the $b$ axis. Two bilayers separated by a van-der-Waals gap
are shown. Top right: local octahedron around a V ion with the corresponding crystal
axes. Lower right: triangular V sublattice projected onto the $ab$ plane and viewed
along $c$. Also shown are the occupied $d_{x^{2}-y^{2}}$ (left) and $d_{xz}$ (right)
orbitals to illustrate the hopping paths provided through overlapping electron
clouds. In reality, neighboring V ions along $a$ are shifted along $c$ (see left
structure).}
\end{figure}

Electronically, VOCl is found to be an insulator. Our photoemission spectroscopy
(PES) data as well as the density functional theory (DFT) calculations confirm the
importance of correlations in this system, and our analysis shows that the
suppression of metallicity is driven by strong correlations. This implies that VOCl
is a multi-orbital Mott-Hubbard system, in line with results from TiOCl, which
displays multi-orbital physics upon electron doping.\cite{Sing09,Zhang09,Craco06}.
On the other hand, the effects of dynamical and spatial fluctuations, which should
be prominently observable in angle-resolved photoemission (ARPES), turn out to be
less important than, e.g., in TiOCl, though not
negligible.\cite{Saha-Dasgupta05,Pisani07}

This paper is structured as follows: in Sec.~\ref{Struc} we give an overview of the
structural, thermodynamic, and magnetic properties of VOCl, in Sec.~\ref{Tech} both
experimental and computational methods are described. Section~\ref{Disc} contains
our results and a comparative discussion of the electronic structure of VOCl at room
temperature and above, determined using (AR)PES as well as DFT calculations within
the generalized gradient approximation including local correlations at the mean
field level (GGA+U). Finally, in Sec.~\ref{Conc} we summarize our results.

\begin{figure}
\includegraphics[width=0.48\textwidth]{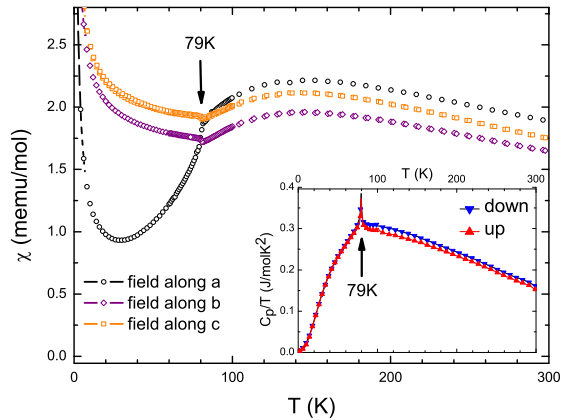}
\caption{\label{Figure2}(Color online) Magnetic susceptibility of VOCl measured with
an external field of 0.1\,T applied along the different crystallographic axes over
an extended temperature range. At $T_{N\acute{e}el}=79$\,K a transition into an
antiferromagnetic phase occurs, accompanied by an anisotropic behavior, evidenced by
the sharp drop if the external field is applied along $a$. The inset shows specific
heat measurements which display a peak anomaly at the same temperature.}
\end{figure}
\section{Structural, thermodynamic, and magnetic properties}\label{Struc}
VOCl crystallizes in an orthorhombic structure of buckled V-O bilayers sandwiched by
Cl layers as shown in Fig.~\ref{Figure1} (space group $Pmmn$; lattice parameters
$a=3.78$\AA, $b=3.30$\AA, $c=7.91$\AA).\cite{Schaefer61} These building blocks are
only weakly coupled through van-der-Waals forces along the crystallographic $c$
direction. Each V ion lies in the center of a strongly distorted octahedron of two
Cl and four O ions. Thus, the degeneracy within the low-lying $t_{2g}$ triplet and
the high-lying $e_{g}$ doublet is completely lifted, as is also the case for
TiOCl.\cite{Hoinkis05,Rueckamp05a,Saha-Dasgupta04} It will be shown below that the
electrons occupy the lowest-lying orbitals of $d_{x^{2}-y^{2}}$ and $d_{xz}$
character in the chosen reference frame ($x=b, y=c, z=a$), which leads to a weaker
electronic anisotropy within the $ab$ plane compared to TiOCl. This enhanced
two-dimensionality is also reflected in the magnetic susceptibility displayed in
Fig.~\ref{Figure2}. While showing an isotropic behavior for the high-temperature
phase, a sharp drop with the external field applied along the crystallographic $a$
axis is observed at around $T_{N\acute{e}el}=79$\,K, while along the $b$ and $c$
axis a slight upturn occurs. This behavior, in conjunction with the anomaly in the
specific heat shown in the inset of Fig.~\ref{Figure2}, indicates a transition to an
antiferromagnetically ordered phase, which has been described previously by a
two-dimensional $S=1$ Heisenberg model on a square lattice with antiferromagnetic
nearest-neighbor interaction.\cite{Wiedenmann83} In this low-T N\'eel phase the
crystal symmetry is reduced from orthorhombic to $c$ axis
monoclinic.\cite{Komarek09,Schoenleber09}

\section{Technical details}\label{Tech}
Single crystals were grown by chemical vapor transport.\cite{Schaefer58,Hoinkis05}
Crystals of typical dimensions $(3\times2\times0.1)$\,mm$^3$ were selected for the
PES measurements. The samples were characterized by x-ray diffraction and magnetic
susceptibility measurements (see Fig.~\ref{Figure2}). For the (AR)PES experiments we
used a Specs PHOIBOS 100 as well as an Omicron EA125 analyzer, monochromatized
Al\,\textsc{K}$_{\alpha}$ radiation for x-ray photoemission spectroscopy (XPS;
$h\nu$=1486.6\,eV) and He\,\textsc{I}$_{\alpha}$ photons ($h\nu$=21.2\,eV) from a
gas discharge lamp for valence band spectroscopy (UPS, ARPES). The total energy
resolution amounted to $\approx$\,700\,meV and 70\,meV for XPS and UPS,
respectively. In ARPES, using the EA125, the angular resolution was about
$1^{\circ}$. Clean surfaces were obtained by \emph{in situ} cleavage using Scotch
tape. Their cleanliness and atomic long-range order were evidenced using XPS and
low-energy electron diffraction (LEED), respectively. Pressure during measurements
was in the low $10^{-10}$\,mbar regime. Due to the strongly insulating nature of
VOCl all measurements have been performed at temperatures of 360\,K or above in
order to minimize charging effects via thermally activated
conductivity.\cite{Hoinkis05} It was found from systematic temperature variations
that for $T>360$\,K the maximum of the V $3d$ peak saturates at $\sim$2.2\,eV below
the chemical potential $\mu_{exp}$ as determined from the Fermi edge of a silver
foil. All measured spectra have been aligned accordingly, and the calculated spectra
have been shifted such that their first order moments correspond to those of the experimental ones.

The DFT calculations where performed with the Full Potential Linearized Augmented
Plane Wave basis (FPLAPW) as implemented in the WIEN2k code~\cite{Wien2k}. We
considered a $k$ mesh of $(15 \times 17 \times 7)$ in the irreducible Brillouin zone
and a RK$_{max}$=7. For the GGA+U calculations we considered the implementation of Dudarev \emph{et al.}~\cite{Dudarev98} in WIEN2k where a unique parameter $U_{eff}=U-J$ is needed as an input. This leaves a certain flexibility for the choices of $U$ and $J$ values provided their difference is $U_{eff}$. NMTO downfolding calculations\cite{Andersen00} were also
performed in order to obtain the important hopping parameters among the V atoms.

\section{Results and discussion}\label{Disc}

\begin{figure}
\includegraphics[width=0.48\textwidth]{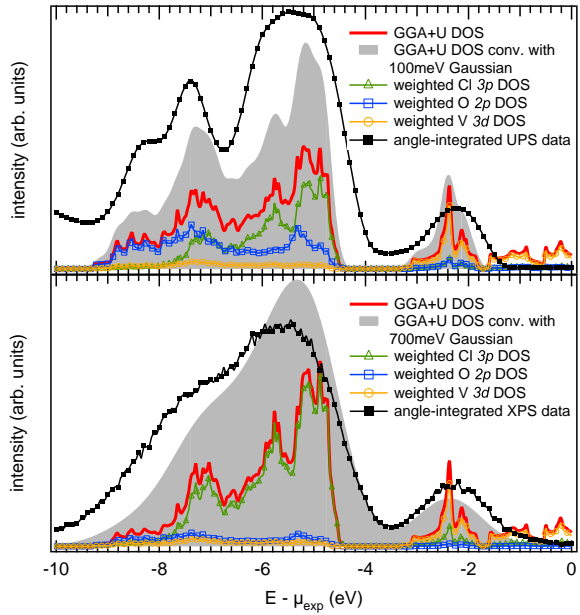}
\caption{\label{Figure3}(Color online) Comparison of calculated partial densities of
states (pDOS) for $U_{eff}=1.3$\,eV with angle-integrated PE spectra (no background
subtraction). The former have been weighted with the respective photo-ionization
cross-sections at the corresponding excitation energies (top: UPS,
He\,\textsc{I}$_{\alpha}$; bottom: XPS, Al\,\textsc{K}$_{\alpha}$). Also shown are
the sums of the weighted pDOS (solid red curves) and their convolutions with a
Gaussian in order to mimic the experimental energy resolution (grey-shaded areas).}
\end{figure}

Figure~\ref{Figure3} presents the angle-integrated UPS and XPS valence band spectra
of VOCl together with the density of states (DOS), calculated for the room
temperature structure within GGA+U for $U_{eff}= U-J =1.3$\,eV where $U$ is the
onsite Coulomb energy and $J$ the exchange integral. Also shown are the partial DOS
(pDOS) for Cl $3p$, O $2p$, and V $3d$ which have been multiplied by calculated
photo-ionization cross-sections~\cite{YehLindau} for UPS (upper panel) and XPS
(lower panel) excitation energies, respectively. In order to mimic the experimental
resolution we convoluted the sums of weighted pDOS (\emph{`GGA+U DOS'}, solid red
curves) with Gaussian functions having a FWHM of $100$\,meV and $700$\,meV for UPS
and XPS, respectively (grey-shaded areas in Fig.~\ref{Figure3}). The experimental
UPS spectrum is representative of an integration in $k$-space around the $\Gamma$
point with a radius of roughly $1/3$ of the Brillouin zone.

\begin{figure}
\includegraphics[width=0.48\textwidth]{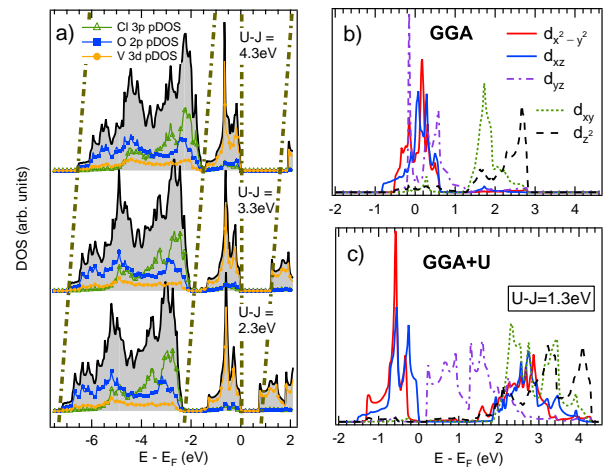}
\caption{\label{Figure4}(Color online) (a) Comparison of the calculated DOS
(grey-shaded areas) using GGA+U with ferromagnetic spin alignment for different
values of $U_{eff}$. While the gap between occupied and unoccupied states decreases
from top to bottom, the separation between the high- and low-binding-energy regions
increases. The widths of these bands, however, remain constant. (b) Orbital-resolved
V $3d$ pDOS calculated using GGA. Three partially filled bands cross the Fermi
energy leading to a metallic state. (c) Same as in (b) but calculated with GGA+U.
Only $d_{x^{2}-y^{2}}$ and $d_{xz}$ are occupied with one electron each, all other
bands are empty. }
\end{figure}

The basic feature observed experimentally, namely the separation of the spectra into
a low-binding-energy part (ca. -2 to -4\,eV) of mainly V $3d$ character and a
high-binding-energy part (ca. -5 to -9\,eV) derived from O $2p$ and Cl $3p$ states is
consistently reproduced in GGA+U.\cite{footnote:bindingenergies} At least for the UPS valence spectrum, the latter
can roughly be further subdivided into a peak of mostly Cl character for
energies of approx. -4 to -6.5\,eV and a structure of mixed O and Cl character below -6.5\,eV.
The changes of shape in the O/Cl bands for different excitation energies are nicely
accounted for by the appropriate cross-sections. While the substructure of these
bands is reproduced remarkably well, the agreement becomes less satisfactory
regarding the shape of the V $3d$ band. This observation will be discussed below in
the context of ARPES data.

An important issue when comparing PES with GGA+U (or LDA+U) calculations is the
choice of the $U_{eff}$ parameters for the GGA+U functional. In our calculations we
treated only the V $3d$ orbitals with the $U_{eff}$ terms. This implies that (i) the
gap between the O/Cl $p$ bands and the V $3d$ bands closer to the Fermi edge as well
as (ii) the charge gap between occupied V $3d$ states and unoccupied ones will
depend on the value of $U_{eff}$. On the other hand, the widths of the different
bands are unaffected and remain constant upon variation of $U_{eff}$. In
Fig.~\ref{Figure4}(a) we show the GGA+U pDOS for three sets of $U_{eff}$ values. For
$U_{eff}=1.3$\,eV the observed PES separation between O/Cl and V occupied bands is
reproduced by the calculations (see Fig.~\ref{Figure3}) although for this value the
gap between the highest occupied and the lowest unoccupied states is underestimated
compared to the gap observed in photoemission as well as to optical absorption
experiments, which reported a charge gap of $\approx$2\,eV.\cite{Venien79,Maule88}
For $U_{eff}=4.3$\,eV the charge gap is basically reproduced but the O/Cl - V gap is
strongly underestimated. Since the latter is fundamentally determined by the
hybridization pattern in VOCl, it has to be captured in the calculations. The
parameter value $U_{eff}=1.3$\,eV is thus the most sensible choice for describing
this system. This value is for instance compatible with  $U=2.3$\,eV and $J=1$\,eV as possible onsite Coulomb and exchange parameters.

\begin{table}
\caption{\label{Table1}Comparison of the $3d$ crystal field splittings (in eV) from
GGA calculations for VOCl and TiOCl, relative to the lowest $3d$ orbital
($d_{x^{2}-y^{2}}$).}
\begin{ruledtabular}
\begin{tabular}{ccccc}
& $d_{xz}$ & $d_{yz}$ & $d_{xy}$ & $d_{z^{2}}$ \\
\hline VOCl & 0.025 & 0.33 & 1.63 & 1.93 \\
\hline TiOCl & 0.241 & 0.46 & 1.54 & 2.08
\end{tabular}
\end{ruledtabular}
\end{table}

The distorted nature of the VO$_4$Cl$_2$ octahedra is reflected in the crystal field
splittings for the V $3d$ orbital energies as shown in Tab.~\ref{Table1}, where
these energies relative to the lowest orbital energy are compared to those of TiOCl.
The values were obtained by considering the first order moment in the DOS GGA
results (see Fig.~\ref{Figure4}(b)). The onsite (i.e. charge neutral) orbital
excitation energies from occupied $d_{x^2-y^2}$ to the different unoccupied orbitals
found in optical absorption experiments are in good agreement with our calculated
values.\cite{Venien79,Maule88,Benckiser08}

We observe that in VOCl the $d_{x^2-y^2}$ and $d_{xz}$ states are almost degenerate
unlike in TiOCl for both GGA and GGA+U. Comparing the two methods, it is important
to note that an insulating state with a finite gap at the Fermi energy is obtained
only when correlations are included. This shows that correlations are the driving
force for the suppression of metallicity in VOCl and that the system is a
(multi-band) Mott insulator. In the case of TiOCl only the $d_{x^{2}-y^{2}}$ orbital
is occupied, defining a dominant one-dimensional hopping path along the $b$
direction.\cite{Saha-Dasgupta04,Seidel03} Nevertheless, from a quantitative
comparison of electronic dispersions in TiOCl and TiOBr as well as from an analysis
of all interaction paths, the importance of two-dimensional correlations for the
electronic properties has been pointed out for these
systems.\cite{Hoinkis07,Zhang08,Aichhorn08}

\begin{figure}
\includegraphics[width=0.48\textwidth]{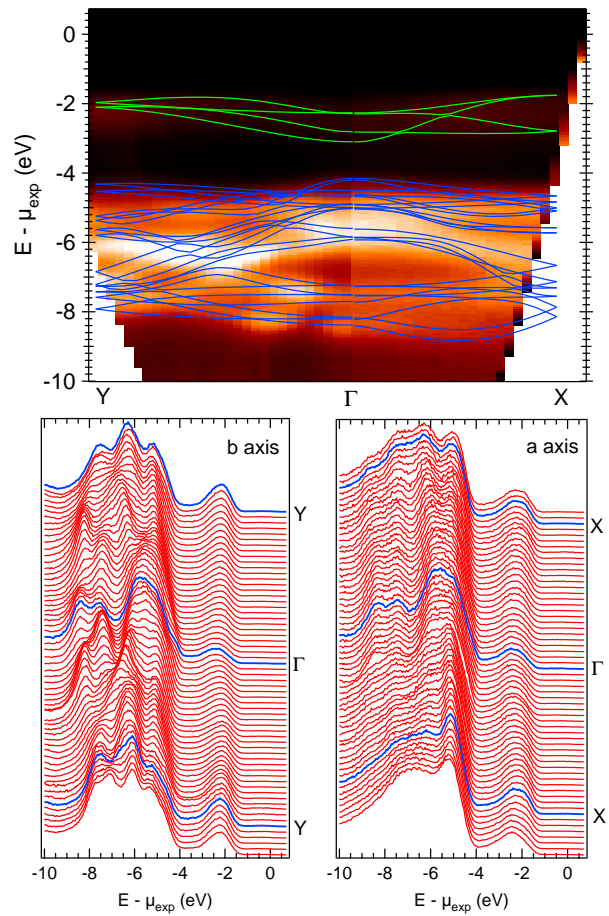}
\caption{\label{Figure5}(Color online) Upper panel: ARPES intensity plots $I(k,E)$
of the valence region along $\Gamma$Y ($b$ axis; left) and $\Gamma$X ($a$ axis;
right) together with band dispersions from GGA+U ($U_{eff} =1.3$\,eV). The latter
have been shifted in energy to account for the effects described in Sec.~\ref{Tech}.
The similarities especially in the high-energy parts (O/Cl bands) are remarkable.
Lower panel: EDCs along Y$\Gamma$Y (left) and X$\Gamma$X (right).}
\end{figure}

From a closer inspection of Fig.~\ref{Figure4}(c) one sees that the total bandwidth
of $d_{xz}$ is larger than that of $d_{x^2-y^2}$, suggesting that strong
antiferromagnetic exchange coupling would be expected along $a$ mainly due to
$d_{xz}$ orbitals. As the latter have a zig-zag overlap along the $a$ axis (see
Fig.~\ref{Figure1}) one can expect that VOCl is intrinsically less one-dimensional
from the electronic structure point of view compared to TiOCl, which is in line with
the observation of a two-dimensional magnetic structure as mentioned
above.\cite{Wiedenmann83,Komarek09} This is verified by calculating energy
differences within GGA+U among different magnetically ordered states and then
mapping the results to a Heisenberg model. With this method we obtain the exchange
integrals between nearest neighbor interacting V ions along $a$ ($J_a$), along $b$
($J_b$) and along the diagonal direction ($J_{ab}$) (see Fig.~\ref{Figure1}). We
performed the calculations for various $U_{eff}$ values and find that $J_a$ is
always larger than $J_b$ and both are antiferromagnetic (e.g. $J_a/J_b=1.1$ at
$U_{eff}=1.3$\,eV and $J_a/J_b=1.6$ at $U_{eff}=3.3$\,eV ) while $J_{ab}$ is smaller
than $J_a$ and $J_b$. However, the nature of the $J_{ab}$ exchange depends on the
choice of the $U_{eff}$ value, e.g., $J_{ab}/J_b=0.31$ at $U_{eff}=1.3$\,eV and
$J_{ab}/J_b=-0.5$ at $U_{eff}=3.3$\,eV. This is probably due to the fact that ferro-
and antiferromagnetic contributions are equally important for the $J_{ab}$ exchange
and with $U_{eff}$ increasing the antiferromagnetic contribution is suppressed,
while the ferromagnetic contribution is barely affected, and $J_{ab}$ is then
ferromagnetic. Nevertheless, the dominant exchanges $J_a$ and $J_b$ are always
antiferromagnetic and their strengths remain comparable for various $U_{eff}$,
indicating that VOCl should be a two-dimensional antiferromagnet at low temperatures
without a preferred one-dimensional hopping path induced by largely different
hopping matrix elements as in TiOCl. Indeed, the NMTO downfolding results identify
the dominant hopping parameters to be of the same order of magnitude in $a$
($0.18$\,eV) and $b$ ($0.13$~\,eV) directions and a factor $\sim 0.6$ smaller in the
$c$ direction. This qualitative trend in dominant hoppings agrees well with the
estimation of the exchange parameters as given above.

\begin{figure}
\includegraphics[width=0.48\textwidth]{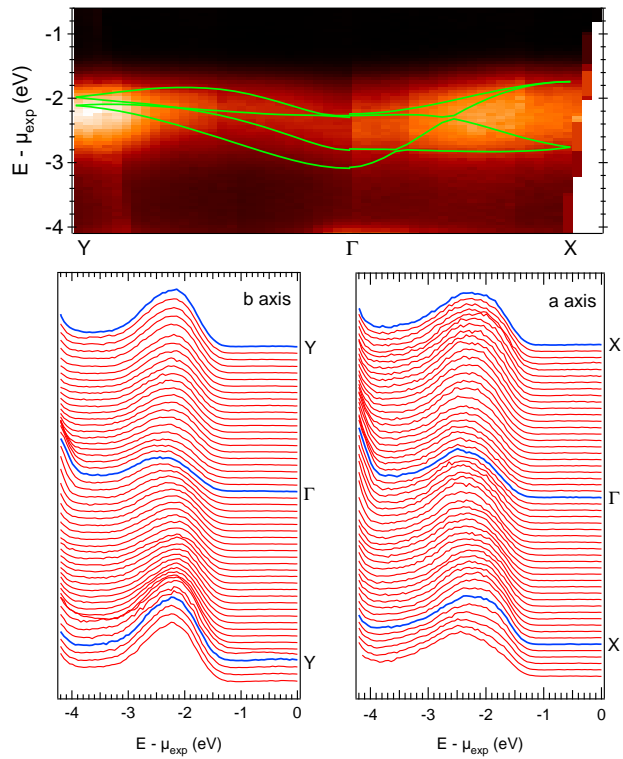}
\caption{\label{Figure6}(Color online)  V $3d$ part of the valence band. Upper
panel: ARPES intensity plots $I(k,E)$ along $\Gamma$Y ($b$ axis; left) and $\Gamma$X
($a$ axis; right) together with band dispersions from GGA+U ($U_{eff} = 1.3$\,eV).
The latter have been shifted in energy to account for the effects described in
Sec.~\ref{Tech}. Lower panel: EDCs along Y$\Gamma$Y (left) and X$\Gamma$X (right).}
\end{figure}

In order to elaborate some more on the dimensionality issue we compare ARPES data
with band structures from GGA+U in Figs.~\ref{Figure5} and ~\ref{Figure6}. First,
Fig.~\ref{Figure5} shows intensity plots of the dispersions of the entire valence
band in the Brillouin zone along $\Gamma$X and $\Gamma$Y corresponding to the
crystallographic $a$ and $b$ axis, respectively, together with the respective energy
distribution curves (EDCs). Also shown is the band structure calculated using GGA+U.

A large number of different bands and rich dispersive features which differ between
the two directions are observed for the O/Cl bands in very good agreement between
theory and experiment. This also indicates the good quality of our single crystals.
In the low-binding-energy region, GGA+U shows a total of four V $3d$ bands since
there are two electrons per V site and two V sites per unit cell. In the ARPES data,
however, the V $3d$ weight appears as a rather broad and weakly dispersing hump
without any sharp features, as can be seen in the blow-up shown in
Fig.~\ref{Figure6}. This behavior was reproducibly observed on many different
samples and is not due to, e.g., a limited experimental resolution. It is rather an
intrinsic property which for several reasons cannot be expected to be fully captured
in our calculations. From a strict many-body view the V $3d$ bands do not have quasi-particle character, but must be interpreted as incoherent weight in the spirit of a Mott-Hubbard scenario. Thus, a description of the detailed line shapes within GGA+U, being an effective single-particle approach which does not account for dynamical and spatial fluctuations, is admittedly intricate. This issue has been extensively studied for the case of TiOCl, where dynamical mean field
theory (DMFT) with a Quantum Monte Carlo (QMC) impurity solver\cite{Saha-Dasgupta05}
in order to take full account of dynamical correlations, as well as the extension to
cluster- rather than single-site calculations,\cite{Saha-Dasgupta07} have led to
considerable improvement. Of course, due to the predominant one-dimensionality in
TiOCl as well as its strong tendency to form singlets (evidenced by the
unconventional spin-dimerized phases at intermediate and low
temperatures\cite{Seidel03}) the influence of such fluctuations is expected to be
stronger than in VOCl. Nevertheless, this issue calls for further investigation in
the present case as well.

Despite these qualifying remarks, we shall give a more detailed analysis in order to compare to the computationally equally complex case of TiOCl. Compared to the latter, the differences of the ARPES dispersions of the V 3d weight between the a and b axis are less pronounced,\cite{Hoinkis05} with the peak maximum appearing at highest binding energies at the $\Gamma$ point and shifting slightly to lower binding energies towards the zone boundary in both cases. A comparison shows that the trend towards lower binding energies as
well as the evolution of the spectral weight distribution are qualitatively in line
with the calculations: large weight is found along $a$ half way between $\Gamma$ and
X, along $b$ right at the Y point, i.e., where bands cross in both cases. The shape
of the peak, which can be seen in the lower panels of Fig.~\ref{Figure6}, reflects
these trends as well: along $b$ it develops from a flat hump to a more distinct,
symmetric peak towards the zone boundary, while along $a$ the slight asymmetry
towards the high-binding-energy side at $\Gamma$ shifts towards the
low-binding-energy side at X. The overall width at the X point appears to be larger than at the Y point, in line with the calculated result that the four GGA+U bands merge at Y while two pairs remain well separated at X. However, a detailed interpretation of this behavior needs further analysis which is beyond the scope of the present work, although we
want to note that the observed asymmetric shape in VOCl is reminiscent of TiOCl
where the asymmetry has been attributed to
correlations.\cite{Hoinkis05,Saha-Dasgupta07}

Overall, the agreement between GGA+U and ARPES for VOCl is better than in the case
of TiOCl. This is probably due to the fact that the influence of dynamical and/or
spatial fluctuation effects in VOCl is smaller than in TiOCl.

\section{Conclusions}\label{Conc}
In conclusion, we have investigated the electronic structure of the low-dimensional
$3d^{2}$ system VOCl using photoemission spectroscopy and GGA+U band structure
calculations. From a comparison to GGA calculations we could show that its
insulating nature is due to strong correlations of the V $3d$ valence electrons
instead of band structure effects, which is the first conclusive evidence for the
Mott character of the insulating state. The issue of choosing $U_{eff}=U-J$ has been
addressed concerning the correct description of the hybridization pattern in favor
of a reasonable magnitude for the charge gap. Calculation of the exchange integrals
shows that VOCl is two-dimensional both electronically and magnetically, and the
obtained antiferromagnetic exchange within the $ab$ plane is robust upon variation
of $U_{eff}$ and in line with experimental observations. Calculated band dispersions
agree well with the ones observed via angle-resolved photoemission spectroscopy for
the high-binding-energy uncorrelated bands, and qualitatively match the V $3d$
derived weight. No considerable in-plane anisotropy within the VO bilayers is
observed in ARPES, supporting the two-dimensional nature of the system. Overall, we
could show that VOCl is a quasi-two-dimensional (as opposed to
quasi-one-dimensional) system and can be well described as a two-band Mott-Hubbard
insulator with $d^2$ ground state configuration.

\begin{acknowledgments}
This work was supported through the Deutsche Forschungsgemeinschaft (DFG) under
Grants CL-124/6-1 and SFB/TRR49 and SFB 484.
\end{acknowledgments}


\end{document}